\documentclass[aps,prl,twocolumn,superscriptaddress,showpacs]{revtex4}

\usepackage{bm}
\usepackage[dvips]{graphicx}

\begin{document}


\title{
 Novel stripe-type charge ordering \\
in the metallic A-type antiferromagnet
 Pr$_{0.5}$Sr$_{0.5}$MnO$_{3}$
}

\date{November 2, 2001}

\author{R. Kajimoto}
\affiliation{Department of Physics, Ochanomizu University, Bunkyo-ku,
Tokyo 112-8610, Japan}

\author{H. Yoshizawa}
\affiliation{Neutron Scattering Laboratory, Institute for Solid State
Physics, University of Tokyo, Tokai, Ibaraki 319-1106, Japan}

\author{Y. Tomioka}
\affiliation{Joint Research Center for Atom Technology (JRCAT), Tsukuba,
Ibaraki 305-8562, Japan}

\author{Y. Tokura}
\affiliation{Joint Research Center for Atom Technology (JRCAT), Tsukuba,
Ibaraki 305-8562, Japan}
\affiliation{Department of Applied Physics, University of Tokyo,
Bunkyo-ku, Tokyo 113-8656, Japan}

\begin{abstract}


We demonstrate that an A-type antiferromagnetic (AFM) state of
Pr$_{0.5}$Sr$_{0.5}$MnO$_{3}$ exhibits a novel charge ordering which
governs the transport property. This charge ordering is stripe-like,
being characterized by a wave vector $\bm{q} \sim (0,0,0.3)$ with very
anisotropic correlation parallel and perpendicular to the stripe direction.
  This charge ordering is specific to the manganites with relatively wide 
one-electron band width ($W$) which often exhibit a \textit{metallic} 
A-type AFM state, and should be strictly distinguished from the CE-type 
checkerboard-like charge ordering which is commonly observed in 
manganites with narrower $W$ such as La$_{1-x}$Ca$_{x}$MnO$_{3}$
 and Pr$_{1-x}$Ca$_{x}$MnO$_{3}$.

\end{abstract}


\maketitle



Perovskite manganites have attracted enormous interests because they
exhibit a colossal magnetoresistance (CMR) effect with hole doping,
where the conductivity shows a significant increase when a ferromagnetic
(FM) state is induced \cite{CMR}.  For the most intensively studied CMR
systems Pr$_{1-x}$Ca$_{x}$MnO$_{3}$ and La$_{1-x}$Ca$_{x}$MnO$_{3}$, a
consensus seems to be reached concerning the mechanism of CMR. In these
systems, a so-called ``CE-type'' charge ordering which appears at $x
\sim 1/2$ is considered to be an essential ingredient. A strong competition
between the FM metallic region and the CE-type charge ordered insulating
 region causes a dramatic change of the transport property.

\begin{figure}[tb]
 \centering \leavevmode
 \includegraphics[width=0.8\hsize]{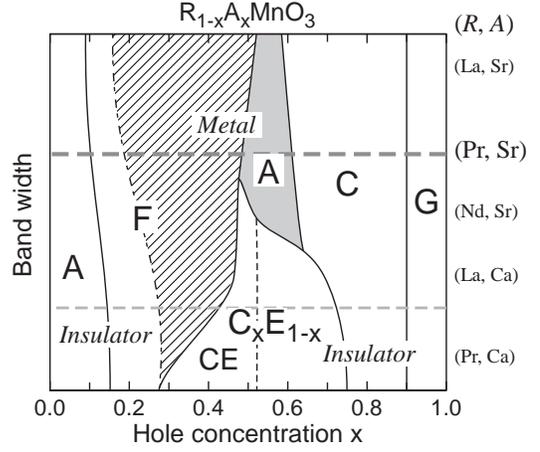}
 \caption{Schematic phase diagram of $R_{1-x}A_{x}$MnO$_{3}$. F
 denotes FM state. A, CE, C, and G denote A-type, CE-type, C-type, and
 G-type AFM states, respectively. C$_{x}$E$_{1-x}$ represents
 incommensurate charge/orbital ordered state \protect\cite{comm1}.}
 \label{fig_diagram}
\end{figure}

One of the important features of these compounds is that these materials
have a relatively narrow one-electron band width ($W$), and exhibits a
wide region of the CE-type charge ordered state on their
hole-concentration versus $T$ phase diagrams \cite{comm1}.  On the other
hand, the CMR phenomenon is not limited to the narrow $W$ manganites,
and is indeed observed in systems with wider $W$ such as
Pr$_{0.5}$Sr$_{0.5}$MnO$_{3}$ \cite{tomioka95_psmo},
Nd$_{1-x}$Sr$_{x}$MnO$_{3}$ with $x \gtrsim 1/2$ \cite{kuwahara99}, and
La$_{2-2x}$Sr$_{1+2x}$Mn$_{2}$O$_{7}$ with $x=0.4$ \cite{moritomo96}.
An important characteristics of these compounds is that all of them show
a \textit{highly conductive} A-type antiferromagnetic (AFM) state, and
some of them even lack the CE-type charge ordering.  The variation of
the phase diagram as a function of $W$ is schematically depicted in
Fig.\ \ref{fig_diagram}.  In contrast to a manganite with a narrow $W$
(indicated by a thin dashed line), a manganite with a relatively wider
$W$ (a thick dashed line) generally shows a following sequence of
spin/charge ordering upon hole doping: insulating A-type AFM
$\rightarrow$ metallic FM $\rightarrow$ \textit{metallic} A-type AFM
$\rightarrow$ insulating C-type, and finally insulating G-type AFM
states.  Clearly, the most important feature here is a lack of the
CE-type spin/charge ordering and the appearance of the \textit{metallic}
A-type AFM \cite{moritomo97,akimoto98}.  In the metallic A-type AFM, a
planar orbital-ordered state with $d(x^2-y^2)$ orbitals is established
around $x=1/2$ \cite{kuwahara99,kaji99}. This orbital ordering mediates
the FM coupling within the orbital-ordered planes in which doped
carriers possess a fairly large mobility \cite{kuwahara99,kawano97}.  At
the same time, the $d(x^2-y^2)$-type orbital ordering favors the AFM
coupling perpendicular to the orbital-ordered planes, and results in an
overall A-type AFM spin state.  The scenario for the CMR phenomenon
based on the CE-type charge ordering is clearly irrelevant in this case,
and another microscopic mechanism ought to be invoked.

\begin{figure}[tb]
 \centering \leavevmode
 \includegraphics[width=0.8\hsize]{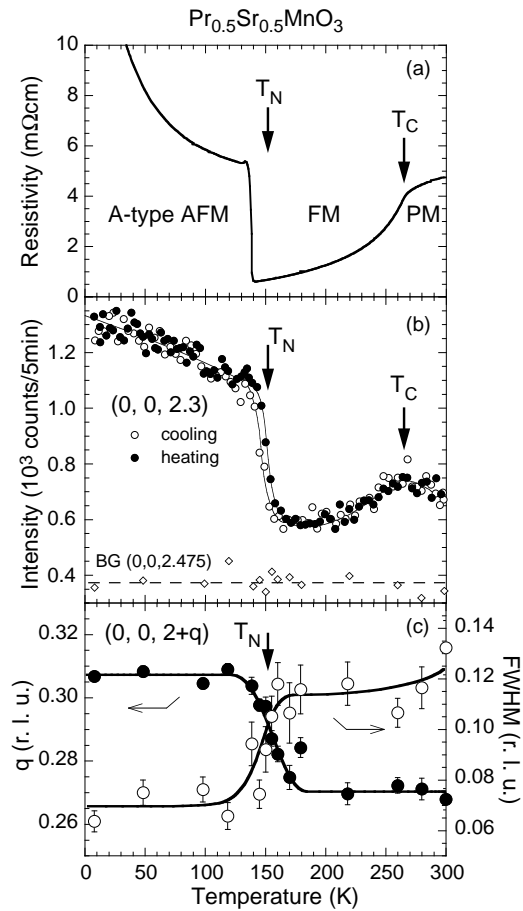}
 \caption{(a) Temperature dependence of the resistivity. (b) Temperature
 dependence of the intensity at $(0,0,2.3)$.  The background (BG)
 intensity measured at $(0,0,2.475)$ is also shown. Open symbols and
 closed symbols denote the data for cooling and for heating,
 respectively. (c) Temperature dependence of the wave vector (closed
 symbols) and the peak width (FWHM) (open symbols) of the charge order
 peak $(0,0,2.3)$.}
 \label{fig_Tdep}
\end{figure}

To unravel this issue, we chose one of the cubic A-type AFM manganites,
Pr$_{0.5}$Sr$_{0.5}$MnO$_{3}$, and performed a detailed neutron
diffraction study.  Pr$_{0.5}$Sr$_{0.5}$MnO$_{3}$ exhibits a first-order
phase transition from a FM metal to an AFM less-conductive state at
$T_{\rm N} \sim 140$ K, accompanied with a structural transition as
shown in Fig.\ \ref{fig_Tdep}(a) \cite{tomioka95_psmo}. It exhibits a
significant MR ($[\rho(0)-\rho(H)]/\rho(H)] > 1000$\%) below $T_{\rm N}$.
  With a high quality
single crystal sample, we found that there exists a novel stripe-type
charge ordering with the modulation vector $q \sim 0.3$~r.l.u.\ (reduced
lattice units) at all $T$.  In what follows, we shall demonstrate that
the quasi-stripe order is intrinsic to the planar $d(x^2-y^2)$-type
orbital ordering, and the transport property in the \textit{metallic}
A-type AFM is, in fact, controlled by the interplay between the
stripe-type charge ordering and the DE interactions.  The present
results provide strong evidence that the physics of the CMR phenomena in
the A-type AFM manganites is fundamentally different from the narrow $W$
manganites where the CE-type charge ordering plays a crucial role.


The single crystal sample well-characterized in the preceding studies
\cite{tomioka95_psmo,kawano97,siozuke} was reinvestigated by neutron
diffraction technique.  An incident neutron momentum $k_{\rm i} = 3.83$
\AA$^{-1}$ and a combination of
40$^{\prime}$-40$^{\prime}$-40$^{\prime}$-80$^{\prime}$ collimators were
utilized at the triple axis spectrometer GPTAS installed at the JRR-3M
reactor in JAERI, Tokai, Japan. Two pyrolytic graphite filters were
placed before and after the sample to suppress higher-order
contaminations.  The sample was mounted in an Al can filled with He gas,
and was attached to the cold head of a closed-cycle helium gas
refrigerator. The temperature was controlled within an accuracy of 0.2
K.

\begin{figure}[tb]
 \centering \leavevmode
 \includegraphics[width=0.75\hsize]{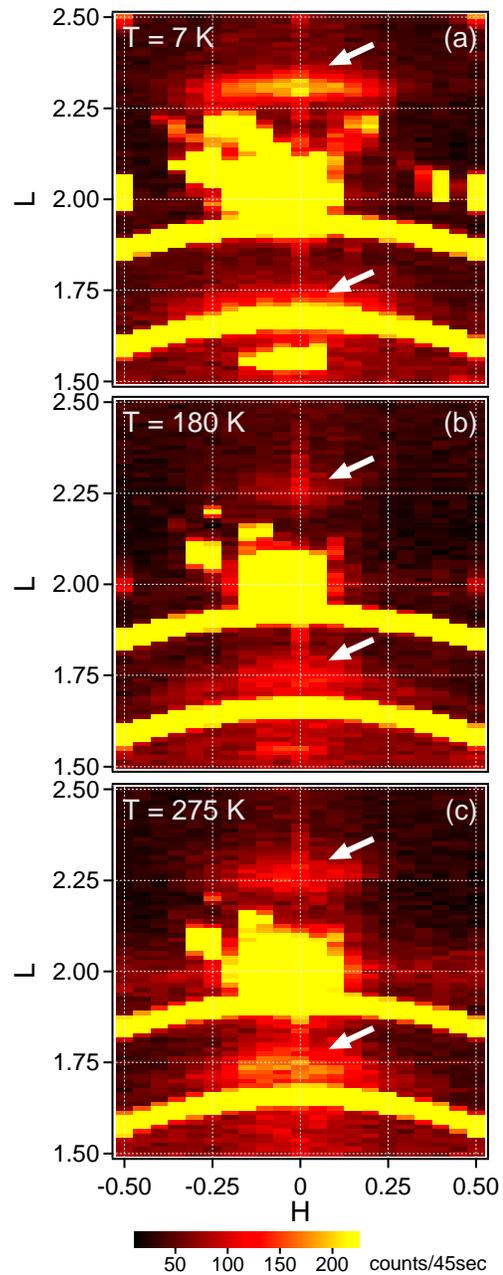}
 \caption{Intensity maps around $(0,0,2)$ at (a) $T = 7$ K, (b) 180 K,
 and (c) 275 K.
 Two ring-shaped scatterings observed at all temperatures
 are due to the Al sample cell. At 7 K, A-type AFM Bragg reflections are
 observed at$(\pm 0.5,0,2)$.}
  \label{fig_map}
\end{figure}

The crystal structure of Pr$_{0.5}$Sr$_{0.5}$MnO$_{3}$ is tetragonal
$I4/mcm$ with $a=b \sim 5.4$~{\AA} and $c/\sqrt{2} \sim 5.5$~{\AA} in
the paramagnetic (PM) and FM states and monoclinic $P2_1/n$ with $a \sim
c \sim 5.4$~{\AA}, $b/\sqrt{2} \sim 5.5$~{\AA}, and $\beta \sim
91^{\circ}$ in the AFM state \cite{kawano97,llobet99}. For simplicity,
we employ the cubic notation with $a \sim 3.8$~{\AA} so that the FM
layers of the A-type AFM structure are perpendicular to the [010]
direction. All the diffraction measurements were carried out on the
$(h,0,l)$ scattering plane to measure the correlations within the FM
layers in the A-type AFM states.


To examine whether Pr$_{0.5}$Sr$_{0.5}$MnO$_{3}$ exhibits a charge
ordering, we surveyed the $(h,0,l)$ scattering plane at selected $T$s.
Figure\ \ref{fig_map} presents a map of the scattering intensities
collected around $(0,0,2)$ in the A-type AFM phase at $T = 7$ K, in the
FM phase at 180 K, and in the PM phase at 275 K, respectively.  The
intense fundamental nuclear Bragg reflection was observed at $(0,0,2)$.
At 7 K, A-type AFM Bragg reflections are observed at $(\pm 0.5,0,2)$.
Two ring-shaped scatterings observed at all $T$s are powder scattering
from the Al sample cell. In addition, twinning and mosaic distributions
of the crystal yield a nuclear reflection near $(-0.1,0,2)$ as well as
an A-type AFM Bragg reflection near $(0,0,1.6)$, but they were slightly
displaced from the commensurate positions due to the small deviations of
the lattice constants of respective domains from the ideal cubic
symmetry.  We also observed the weak CE-type superlattice reflections
appearing at $\bm{Q} \sim (\pm 0.2, 0, 2.2)$ in the AFM phase [Fig.\
\ref{fig_map}(a)].  In the present case, however, the CE-type ordered
region has practically no influence on the transport property, because
its volume fraction is negligibly small, being consistent with the lack
of the CE-type AFM reflections in the previous powder neutron
diffraction study \cite{kawano97}.  We note that the coexistence of the
A-type metallic AFM region and the parasitic CE-type insulating region
was frequently observed in the manganites with $x \sim 1/2$
\cite{kaji99,moritomo98,kubota99}.

The important new results in Fig.\ \ref{fig_map} are anisotropic diffuse
scatterings indicated by white arrows. They are centered at $\bm{Q} \sim
(0,0,2 \pm 0.3)$, and are elongated towards the [100] direction.  The
anisotropy of their profiles indicates that the correlation length along
$\bm{Q}$ is much longer than that perpendicular to $\bm{Q}$.  Although
the diffuse scattering becomes weaker in the FM and PM states, it
remains finite, but the position of the signal shifts slightly towards
$\bm{Q} \sim (0,0,2 \pm 0.25)$.  Because the similar scatterings are
observed at large $Q$ but less clear at small $Q$, the origin of the
scatterings is attributable to lattice modulations.  Due to the twinning
domains with propagation vectors $\bm{q} =(0,0,0.3)$ or $(0.3,0,0)$, the
similar scattering is expected at $(\pm 0.3,0,2)$ in Fig.\
\ref{fig_map}(a), but we found no signal at corresponding
positions. This suggests that the lattice modulations consist of a
longitudinal component, because the scattering cross section has a term
$|\bm{Q} \cdot \bm{\eta}|^2$ where $\bm{\eta}$ represents a displacement
vector of constituent ions. A similar feature was also observed in the
stripe-like charge ordering in a two-dimensional (2D) A-type AFM
manganite La$_{2-2x}$Sr$_{1+2x}$Mn$_{2}$O$_7$ \cite{kubota00,doloc99}.

In order to confirm that the newly found diffuse scattering is
originated from the charge ordering, its influence on the transport
property was examined by the $T$ dependences of the intensity, the peak
width, and the wave vector of the diffuse scattering.  We found a strong
correlation between the resistivity and these quantities, which
establishes that the observed diffuse scattering indeed arises from the
charge ordering.  In Figs.\ \ref{fig_Tdep}(a) and (b) are plotted the
$T$ dependence of the resistivity and that of the diffuse intensity at
$(0,0,2.3)$, respectively.
Fig.\ \ref{fig_Tdep}(c) shows
the $T$ dependence of the peak position (the modulation wave vector) and
the peak width (FWHM) measured along the longitudinal direction
(perpendicular to the stripes). In the A-type AFM state for $T < T_{\rm
N}$, the correlation of the charge ordering is well-developed. The
diffuse peak has large intensity, and its width is quite sharp, although
it is much broader than the instrumental resolution ($\sim
0.03$~r.l.u.). In this phase, the resistivity exhibits a steep increase
with lowering $T$.  At $T_{\rm N}$, the system undergoes the first order
transition.  The intensity of the charge ordering and the resistivity
show sudden decreases at $T_{\rm N}$.  In addition, the correlation
length of the charge ordering decreases, while its peak position shifts
to a smaller wave vector on crossing $T_{\rm N}$.  Upon raising $T$,
however, the intensity of the charge ordering gradually increases, and
the resistivity recovers the metallic behavior.  Around $T_{\rm C}$, the
resistivity as well as the intensity of the charge ordering shows a cusp
and enters the PM state.  Note that the novel diffuse scattering
subsists in the FM as well as PM states in
Pr$_{0.5}$Sr$_{0.5}$MnO$_{3}$.

These results clearly indicate that the newly-discovered charge ordering
is intrinsic to the A-type AFM Pr$_{0.5}$Sr$_{0.5}$MnO$_{3}$ with
$d(x^2-y^2)$ orbital ordering.  In particular, the transport property of
Pr$_{0.5}$Sr$_{0.5}$MnO$_{3}$ is controlled by the interplay between the
DE interactions and the novel charge ordering which is embedded in the
MnO$_2$ planes with the $d(x^2-y^2)$ orbital ordering.  Since a similar
charge ordering was observed in the 2D A-type AFM manganites, the
existence of such charge ordering must be independent to the spatial
dimensionality of the manganites, and is one of the key features of the
wide $W$ manganites which are accompanied with the
\textit{$d(x^2-y^2)$-type orbital ordering}.

\begin{figure}[tb]
 \centering
 \includegraphics[width=0.8\hsize]{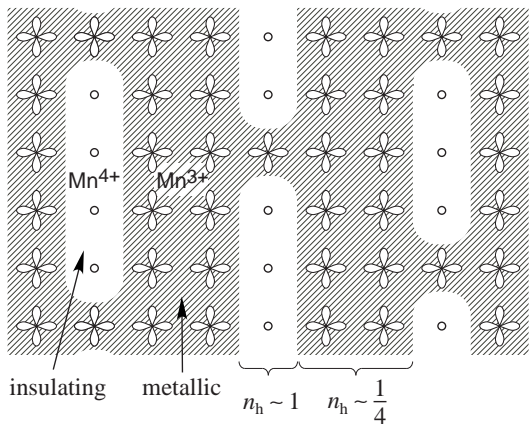}
 \caption{Schematic illustration of the charge order and orbital order
 with $q = 1/3$~r.l.u. Cloverleaf symbols represent the $d(x^2-y^2)$
 orbitals. $n_{\rm h}$ denotes the hole concentration within the
 Mn$^{4+}$ stripe or the Mn$^{3+}$-like matrix.}
 \label{fig_model}
\end{figure}

We propose that the simplest model of the charge ordering which is
compatible with these observations may be a stripe-like object drawn in
Fig.\ \ref{fig_model}. In this model, Mn$^{4+}$ ions segregate within
the metallic matrix of Mn$^{3+}$-like sites with $d(x^2-y^2)$ orbitals,
and form stripe-like objects along the Mn-O-Mn bond direction.  The
Mn$^{4+}$ stripes are insulating, and they block the hopping of the
$e_g$ electrons, while the Mn$^{3+}$-like matrix is metallic.
In this sense, the conductivity of the A-type AFM system is controlled by the
number of stripe-like objects within the $d(x^2-y^2)$-type
orbital-ordered matrices.  Thus, the A-type AFM system
can be highly conductive, and in this sense, it is distincly
different from the insulating CE-type charge ordering.

The observed stripe-like ordering has some distinct features.  First of
all, the amplitude of the wave vector $q \sim 0.3$ r.l.u.\ is far apart
from its nominal hole concentration $x = 1/2$. For the model with $q =
1/3$ r.l.u.\ depicted in Fig.\ \ref{fig_model}, the hole concentration
of the Mn$^{4+}$ stripe is fixed to 1, while that of two Mn$^{3+}$ lines
be 1/4, resulting in the overall concentration $x = 1/2$.  The large
discrepancy between the hole concentration and the amplitude of the wave
vector is also observed in 2D manganites
La$_{2-2x}$Sr$_{1+2x}$Mn$_{2}$O$_{7}$ where the wave vector is
effectively fixed at $q \sim 1/3$ for a wide region of the hole
concentration \cite{kubota00}.  It should be noted, however, that the
wave vector of the stripe-like diffuse scattering in
Pr$_{0.5}$Sr$_{0.5}$MnO$_{3}$ is slightly shifted through the transition
from the A-type AFM state to the FM state as seen in Fig.\
\ref{fig_Tdep}(c).  This behavior indicates that the mechanism of the
formation of the stripe-type charge ordering in wide $W$ manganites is
succeptible to the change of the spin structure and concomitant change
of the electronic state.

An effectively fixed wave vector at $q \sim 1/3$ suggests that the
relative distance between such stripes remains almost the same for a
wide region of the hole concentration by tuning the hole concentrations
within stripes and the Mn$^{3+}$-like matrix regions.  This
interpretation is consistent with the observed strong anisotropy of the
diffuse scattering.  The striking asymmetry of the profiles of the
diffuse scattering was observed as shown in Fig.\ \ref{fig_map}. The
correlation length $\xi_{\parallel}$ parallel to the stripe is much
shorter than $\xi_{\perp}$ perpendicular to the stripe. From the width
of the profile along the [100] direction, $\xi_{\parallel}$ is estimated
to be $4a \sim 5a$.  These results are consistently explained by the
proposed stripe picture.  In particular, $e_g$ electrons enter every $4
\sim 5$ sites of the line of Mn$^{4+}$ ions as depicted in
Fig. \ref{fig_model}.  The tuning of the hole concentration causes the
deviation of the hole density within stripes to be $1-\delta$, while
that of the Mn$^{3+}$-like matrix regions $1/4+\delta/2$, respectively.
For a finite $\delta$, some excess electrons intervene Mn$^{4+}$
stripes, while some holes are accommodated within the Mn$^{3+}$-like
$d(x^2-y^2)$-type orbital-ordered matrix.  Consequently, the stripe
ordering consist of a number of short segments as shown in Fig.\
\ref{fig_model}.

One more important feature of the present stripe-like charge ordering is
that the charge stripes are parallel to the Mn-O bonds in contrast to
the CE- and C$_{x}$E$_{1-x}$-type charge ordering \cite{incomme}.  The
direction of the stripes can be explained by the large mobility within
the orbital-ordered planes. The DE mechanism mediated by the hoping of
holes is a dominant interaction within the metallic orbital-ordered
planes, while the superexchange (SE) interactions are dominant within
stripes and at the boundaries between Mn$^{4+}$ stripes and metallic
regions. In the case of the ``parallel'' charge order, the number of the
Mn-O-Mn bonds mediating the DE interactions is much larger than that in
the case of the ``diagonal'' charge order. When $W$ is sufficiently
large, the DE interaction overwhelms the SE interaction, and the
parallel charge order is favored by maximizing the number of the DE
interactions. On the contrary, when $W$ is small, the SE interactions
dominate, and favor the diagonal charge order known as the CE- and
C$_{x}$E$_{1-x}$-type \cite{mizokawa98}. We would like to point out
that a similar relation between the charge mobility and the direction of
stripes holds in the well-known stripe ordered systems, {\it i.e.} 
high-$T_c$ cuprates \cite{lsco} and isomorphic nickelates \cite{lsno}. 
In the metallic cuprates, the stripes are ``parallel'', while they are
``diagonal'' in the insulating nickelates.  Note that the $d(x^2-y^2)$
orbital state is also common to both A-type AFM manganites and
high-$T_c$ cuprates: the stripes in both \textit{metallic} materials
are formed in the matrix of $d(x^2-y^2)$ orbital states.


To summarize, a neutron diffraction study was performed on a single
crystal of Pr$_{0.5}$Sr$_{0.5}$MnO$_{3}$. We found a very anisotropic
charge ordering with wave vector $\bm{q} \sim (0,0,0.3)$. The diffuse
scattering is consistent with the stripe-like charge ordering.  This
novel charge ordering exists in the manganites with a wide one-electron
band width and with the $d(x^2-y^2)$-type orbital ordering regardless of
its spatial dimensionality.  The transport property is controlled
through the competition between the novel stripe-like charge ordering
and the DE interactions.  The CMR phenomenon in these materials must be
understood on the basis of this \textit{``stripe-like charge
ordering''}.

\begin{acknowledgments}

We thank H. Kawano-Furukawa for valuable discussions.  This work was
supported by a Grand-In-Aid for Scientific Research from
MEXT, Japan and by NEDO of Japan.

\end{acknowledgments}


\begin{thebibliography}{99}
\bibitem{comm1} The incommensurate charge and orbital order observed at
	 $x > 1/2$~\cite{incomme} can be regarded as an extension of the
	 CE-type charge/orbital order, and we represent it by
	 C$_{x}$E$_{1-x}$ in Fig.\ \ref{fig_diagram} following Ref.\
	 \cite{dagotto01}.
\end{thebibliography}
\end{document}